\documentclass[a4paper,twoside,10pt]{article}
\usepackage{graphicx,publaob}

\def\papertype{Invited Review}
\begin{document}

\title{INVESTIGATION OF ACTIVE GALACTIC NUCLEI  IN TIME DOMAIN ERA}

\authors{D. Ili\' c$^1$, A. Kova\v cevi\'c$^1$   \lowercase{and}  L. \v C. Popovi\'c$^{1,2}$}

\address{$^1$Department of Astronomy, Faculty of Mathematics, University of Belgrade, Studentski trg 16, Belgrade, Serbia}
\Email{dilic}{matf.bg.ac}{rs} \Email{andjelka}{matf.bg.ac}{rs}
\address{$^2$Astronomical Observatory, Volgina 7, Belgrade, Serbia}
\Email{lpopovic}{aob}{rs}



\markboth{INVESTIGATION OF AGNs IN TIME-DOMAIN ERA}{D. ILI\'C, A. KOVA\v CEVI\'C, L. \v C. POPOVI\'C}

\abstract{The perfect case for time-domain investigations are active galactic nuclei (AGNs) since they are luminous objects that show strong variability. Key result from the studies of AGNs variability is the estimated mass of a supermassive black hole (SMBH), which resides in the center of an AGN. Moreover, the spectral variability of AGN can be used to study the structure and physics of the broad line region, which in general can be hardly directly observed. Here we review the current status of AGNs variability investigations in Serbia, in the perspectives of the present and future monitoring campaigns.
}

\section{INTRODUCTION}

With the advances of technology and robotization of telescopes, we have entered the era of massive astronomical time-domain observations, with several missions already running and producing amazing results (e.g. Djorgovski et al. 2016). This is the case also for the investigations of active galactic nuclei (AGNs).
The Sloan Digital Sky Survey (SDSS) is one of the most famous and successful example, which is currently releasing its sixteenth data release containing images and spectra of millions of objects, as well as providing a Time Domain Spectroscopic Survey of 131,000 quasars (e.g. Ahumada et al. 2020). In the next decade, we expect to have comprehensive monitoring campaigns, out of which the Vera Rubin Observatory Legacy Survey of Space and Time (LSST) with its very high cadence 10-year photometric survey seems to be very promising for the quasar investigations (Ivezi\'c et al. 2019).  LSST will provide millions of quasar photometric light curves, which will facilitate better understanding of the physical processes that power quasars and their variability (Brandt et al., 2018). LSST will be complemented with several spectral surveys, and one promising example is the Maunakea Spectroscopic Explorer (MSE) that is an 11m aperture telescope (MSE Science Team 2019). This survey will cover both the time and spectral domain, including also the infrared band, which is important for the high-redshift quasar studies. The MSE will lead the world in multi-object spectroscopy, with its unique capability to study up to 4,000 astronomical objects at once. It will provide spectral light curves of the largest number of quasars, which analysis will enable mapping of the inner regions in AGNs (Shen et al. 2019).

This motivates us to do investigations of AGNs using optical spectroscopy in time-domain, of which recent results we give an overview in this Proceedings. We cover topics from: i)  long-term monitoring campaign of the sample of near-by AGNs, ii) studies of oscillations in AGN light curves in search for periodic signals, iii) studies of the optically extremely variable AGNs. The findings of our research project are put in the concept of applications and impact to future large surveys, especially the LSST. These investigations are done at the Astronomical Observatory in Belgrade and the Department of Astronomy of the Faculty of Mathematics, University of Belgrade, mainly through the project of "Astrophysical spectroscopy of extragalactic objects", 176001, supported by the Ministry of Education, Science and Technological Development of the Republic of Serbia, which was financed by the end of 2019. 
This paper is organized as follows: in Sec. 2 we shortly describe the theoretical background of AGN structure, and give the current overview of time-domain investigation of AGNs, in Sec. 3 we list some of the major results of our findings, in Sec. 4 put them in concept of future surveys, and in Sec. 5 a short summary is given.

\section{OVERVIEW OF AGNs AND TIME-DOMAIN INVESTIGATIONS}

\subsection{Active galactic nuclei}

AGNs are nested in the center of only a small fraction of all galaxies (Netzer 2006). However, it is widely believed that every larger galaxy has undergone a phase of activity, when its supermassive black hole (SMBH) was actively accreting matter from the surroundings. This process of accretion is responsible for the production of powerful and energetic continuum emission, which produces a variety of different phenomena, such as the production of relativistic radio-jets, disk-winds, or ionized gaseous regions (for a textbook on AGNs, see e.g. Netzer 2006). The broad optical emission lines originate in the broad line region (BLR), which is photoionized by the continuum emission from the accretion disk. The accretion disk and BLR are sometimes "hidden" within the larger dusty region, which is a dominant source of obscuration of optical emission in some type of AGN. Dust is mostly located in equatorial plane and is usually referred as a dusty "torus", but recent high-resolution observations showed that the polar dust is also present (H{\"o}nig \& Kishimoto 2017, Stalevski et al. 2017, 2019).

Quasars are classified in many ways, depending on their observed properties. One division is on type 1 and type 2 AGNs (or unobscured and obscured AGNs) which is based on the presence or absence of optical broad emission lines, respectively.  The main reason for this division is to be the viewing angle, i.e. the angle how the AGN axis of symmetry (i.e. the accretion disk axis) is inclined with respect to the observer. If we would be looking at the AGN closer to the AGN axis, we would be seeing inside the dusty region, thus detecting the unobscured accretion disk and BLR, i.e. broad emission lines (Antonucci \& Miller 1985,  Urry \& Padovani 1995). When the viewing angle is larger, we would be seeing the central parts covered by the dusty region, and the continuum and broad emission lines would be obscured, as it is in the case of type 2 AGNs (Hickox \& Alexander, 2018).

We are especially focused on the BLR investigations through the analysis of the optical spectroscopy. Primarily because one can estimate the SMBH mass from the dynamics of the BLR gas, which is gravitationally bound to a SMBH. For this we need to measure the BLR size, which is measured through the reverberation mapping (RM, see e.g., Blandford \& McKee 1982, Gaskell \& Sparke 1986). The RM measures a characteristic size of the BLR from the time lag (i.e. light echo) between variability of the continuum emission, which powers the BLR, and the delayed response of the BLR emission. The RM was applied to a limited sample of AGNs, even if the SDSS survey results  are considered (e.g. Shen et al. 2019). However, it has produced the so-called “single-epoch virial BH mass estimators” (e.g., Vestergaard \& Peterson 2006)  that provides an empirical SMBH mass estimate using the single-epoch spectroscopy and the relation between the BLR radius and continuum luminosity (so-called radius-luminosity scaling relation, see e.g., Bentz et al. 2009). This method has been widely applied to AGNs at different redshifts and luminosities.

We still do not have a complete physical picture of the BLR which is known to be very complex (Sulentic et al. 2000), and this can directly affect the accuracy of the SMBH mass estimates (as discussed in e.g. Mej{\'\i}a-Restrepo et al. 2018).
There are many opened question about the geometry and kinematics, e.g. is the BLR truly virialized to the SMBH (Joni\'c et al. 2016, Popovic et al. 2019), do we have a Keplerian motion, outflows or inflows (e.g. Wang et al. 2017), what is the inclination of the moving gas (e.g. Mej{\'\i}a-Restrepo et al. 2018, Afanasiev et al. 2019). The physical properties (e.g. gas temperature, density, ionization parameter) of the emitting plasma in the BLR are still not well constrained (Ili\'c et al. 2012). Currently the BLR is not spatially resolved, except with the state-of the-art interferometry such as GRAVITY (Gravity Collaboration, Sturm et al. 2018, Amorim et al. 2020, 2021, { Wang et al. 2020, Songsheng et al. 2021}) or the proposed upgrade GRAVITY+ (Eisenhauer 2019)\footnote{https://www.mpe.mpg.de/7480772/GRAVITYplus\_WhitePaper.pdf}.  Thus the spectroscopy is still important tool to observe and study the BLR.

\subsection{Time-domain astronomy of AGNs}

The International Astronomical Union (IAU) has very recently recognized that the time-domain investigations are an important aspect of the modern research and has founded a Working Group on Time Domain Astronomy in 2015 (see e.g. their 2019 annual report\footnote{https://www.iau.org/static/science/scientific\_bodies/working\_groups/260/wg-tda-annual-report-2019.pdf}).  Time domain astronomy studies the transient Universe and covers objects from the near-by Solar system, through variable starts, to galaxies at cosmological distances, i.e. quasars. As already stated, this is strongly driven by current and coming large time domain surveys, such as the Catalina Real-Time Transient Survey (CRTS, Drake et al. 2009), Zwicky Transient Facility (ZTF, Ofek et al. 2020), and the coming LSST. 

Quasars are in the focus of the time-domain investigations (Graham, 2019), since one of the most peculiar observational feature of quasars is its variability, which can be on the orders of hours to months, depending on the wavelength band (see e.g. Peterson, 2001). Especially, the broad emission lines can strongly change both the flux and profile. The flux variations were used for the RM studies to "resolve" the BLR in time-domain and measure the size of the BLR (e.g. Peterson et al. 2004), and consequently the SMBH mass (see review Popovi\'c 2020), whereas the line profile variability can be used to constrain the geometry of the BLR (e.g. Popovi\'c et al. 2011).
Particularly important aspect of time-domain photometric observation of AGNs, is the search for periodic signal which could lead to the detection of close binary SMBH systems (see e.g. with CRTS, Graham et al. 2017) and possible future targets for detection of gravitational waves (for a  review see Popovi\'c 2012, and recent works Kova\v cevi\'c et al., 2019, 2020b). One particular class of AGNs are so-called changing-look AGNs (CL AGNs), which show extreme variability of emission line intensities and profiles, with sometimes almost complete disappearance and reappearance of the broad component in some emission lines (e.g. Denney et al. 2014, Oknyansky et al. 2017). Physical processes causing such dramatic change are still unknown (e.g. Ili\'c et al. 2020), but this could be also a signature of a presence of close binary SMBH (Wang \& Bon 2020). So far, surveys have discovered only a small fraction of changing-look candidates (e.g., Runco et al. 2016, MacLeod et al. 2016), but future surveys like the LSST will be able to discover many or even show that this phase is not so rare.


\begin{figure}
\center
\includegraphics[width=11cm,height=18.6cm]{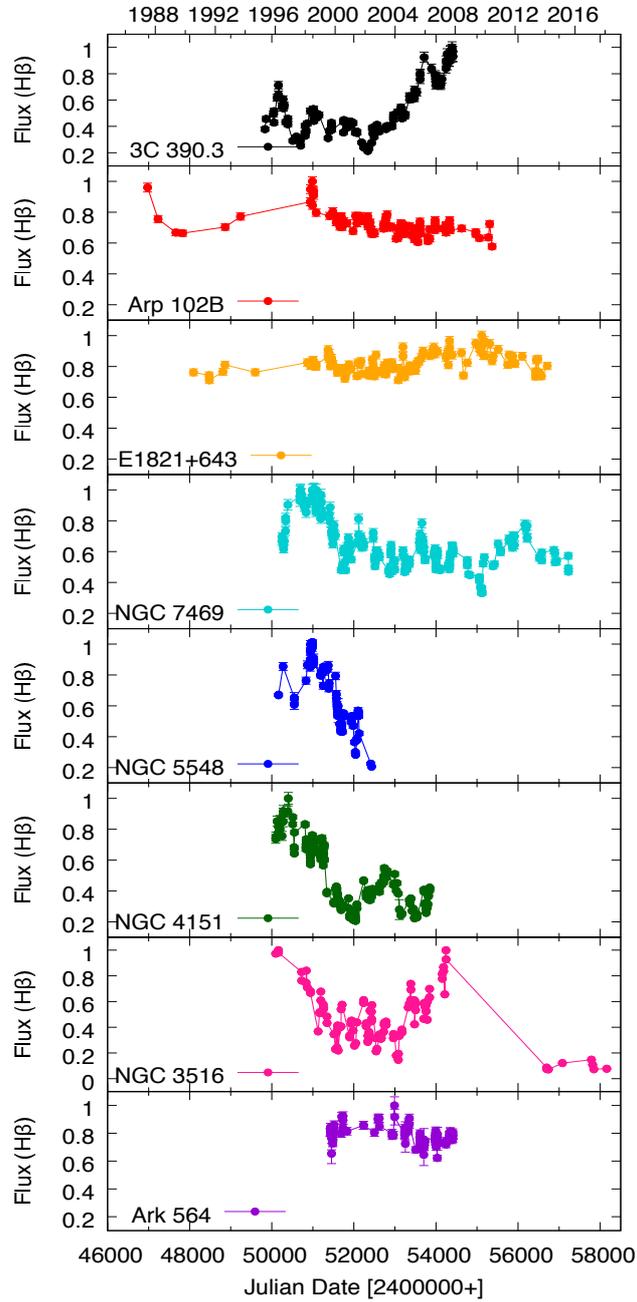}
\caption{H$\beta$ light curves of the sample of eight type 1 AGNs from the LoTeRM campaign (only data presented in the first data release papers are shown). Different sub-types of type 1 AGNs are shown, which exhibit different level of variability and optical spectral lines behavior. Light curves are normalized to the maximal flux.}
\end{figure}

\section{RESULTS OF OUR STUDIES OF AGNs VARIABILITY}

\subsection{Long Term Reverberation Mapping (LoTeRM) campaign}

The long term spectroscopic AGNs monitoring program (Long Term Reverberation Mapping - LoTeRM) was a programme initiated by late A.I. Shapovalova with the aim to study the structure of BLR and estimate the mass of the SMBH. LoTeRM has been running for the last $\sim$30 years collecting data from several world-wide optical telescopes (see Shapovalova et al. 2009, Ili\'c et al. 2017, for reviews). 
The LoTeRM sample included type 1 AGNs of different variability properties (e.g., low or high level of variability, in the range of 5-40\%), and different spectral line profiles (e.g., double-peaked profiles or asymmetric profiles). So far the data have been published for the following sub-types: Seyfert 1 galaxies (NGC 5548, NGC 4151, NGC 7469), Narrow-line Seyfert 1 galaxy - NLSy 1 (Ark 564), double-peaked line radio loud (3C 390.3) and radio quiet (Arp 102B) galaxy, luminous quasar (E1821+643), a SMBH binary candidate, and a CL AGN (NGC 3516). Each object has been first presented and analyzed separately in a series of papers (see release paper Shapovalova et al. 2001, 2004, 2008, 2010a, 2012, 2013, 2016, 2017, 2019). Spectral data were all uniformly reduced and analyzed, creating a unique sample of homogeneous long-term light curves of different properties (Figure 1), making it a valuable set for studies of AGN variability. As a part of the Shapovalova et al. monitoring campaign, {in Bon at al. (2016) additional observations of NGC 5548 covering 2003--2013 are given. In the same paper an important historical light-curve of NGC 5548 which covers more than 40 years of observations was constructed}.
Some of the results of LoTeRM were included in the AGNs Black Hole Mass Database (Bentz \& Katz  2015)\footnote{http://www.astro.gsu.edu/AGNmass/}.

The LoTeRM project is currently expanding to include other telescope, such as the 1.4m Milankovic telescope at Astronomical Station Vidojevica, Serbia, and 1.5m telescope of the Observatory of Sierra Nevada, Granada, Spain. The Milankovic telescope is for the moment only equipped for the photometric monitoring, which can be used in combination with the spectral data or carefully selected narrow/mid-band filter for the RM applications (Malygin et al. 2020). This was used in case of NGC 4395, which is a bulgeless dwarf galaxy with one of the smallest SMBH of only 10,000 solar masses (Woo et al. 2019).  This SMBH mass was confirmed with the international intranight RM campaign done by more than a dozen of telescope across the world including the 1.4m Milankovic telescope (Cho et al. 2020). The observed time delays of the BLR were only $\sim$1-2 hours, and it was important to show that this case of very low-luminosity AGN follows the radius-luminosity scaling relations (Bentz et al. 2013).

\subsection{Oscillatory patterns in AGN light curves }

We used this sample of AGNs to investigate the behavior of long-term light curves, and search for periodicity or patterns which could signify different underlying dynamical processes. In case of 3C 390.3, it was shown that quasi-periodical oscillations may be present in the continuum and H$\beta$ light curves (Shapovalova et al. 2010). Possibly, the outbursts seen in the H$\beta$ line could be explained by successive occurrences of two bright orbiting spots in the accretion disk (Jovanovi\'c et al. 2010). Careful investigation of the repeating variability pattern in complex broad emission line profiles, resulted with a first paper on the spectroscopically discovered binary orbit of SMBH binary system candidate in NGC 4151 (Bon et al. 2012). Later it was shown that a binary SMBH system may be possibly present in another well-known object NGC 5548 { (e.g. Li et al. 2016), since significant periodicity in light and radial velocity curves in NGC 5548 were detected (Bon et al. 2016)}.

Furthermore, a novel hybrid method to search for periodic oscillatory patterns was developed -  2DHybrid method (see Kova\v cevi\'c et al. 2018, for details). The 2DHybrid combines two techniques continuous wavelet transform and correlation coefficients, to analyze the Gaussian-processed AGN light curves. We have applied the 2DHybrid to our LoTeRM sample (Kova\v cevi\'c et al. 2018, 2020a) finding periodic variations in most of them, but of different origin. In case of the two double-peaked line objects 3C 390.3 and Arp 102B, for which  we found a good evidence that they have qualitative different dynamics (Kova\v cevi\'c et al. 2018). 

With the 2DHybrid method we probed the light curve of the famous case of a binary SMBH candidate PG1302-102, which photometric light curve showed clear periodic behaviour (Graham et al., 2015). More recent optical observations showed a disturbed light curve of PG1302-102, which lead to less significant periodicity detection (Liu et al. 2018). However,  our hybrid method for periodicity detection was able to identify the periodic signal even in light curve with larger fluctuations (Kova\v cevi\'c et al. 2019), and proposing a model of a binary system in which a perturbation in the accretion disk of a more massive component is present (Simi\'c \& Popovi\'c 2016). This model of a binary system has been further developed to include a  SMBBH system which considers that both SMBH have their accretion disc and
BLR regions, and are both surrounded by a common circumbinary BLR (Popovi\'c et al. 2021).

\subsection{Extremely variable AGNs}

Long-term monitoring of AGNs is particularly important to understand the trend of their variability and capture possible extreme events in the optical band, such as the changing look transition, i.e. the disappearance of the broad component in some emission lines. Moreover, it could be that the CL phenomenon is a common phase, and that it could be detected in each strongly variable AGN if constantly observed (see discussion in Oknyansky et al 2017). The CL transition could be caused by the intrinsic changes of accretion disk, obscuration of the BLR, tidal disruption events, or even presence of the close binary SMBH (Wang \& Bon 2020). Therefore, understating these object could help shedding light on understanding the AGN structure and evolution. 

Our sample contains several cases of extremely variable objects in the optical band (e.g. NGC 4151, NGC 5548), and the CL state for NGC 5548 was reported in Shapovalova et al. (2004). Another example of CL AGN is a Seyfert galaxy NGC 3516, for which is known to have strong optical variability and has changed its type in the past (Andrillat \& Souffrin 1968). For NGC 3516, the LoTeRM campaign collected a 22 years of data, from 1996 to 2018, and we found strong variability in the continuum and broad lines (Shapovalova et al. 2019). The minimum of activity was observed in 2014, when the broad lines almost disappeared. In 2017, the object was still in the low state, but broad lines started to appear (see Figure 2 in Shapovalova et al. 2019). Recent short-term intensive optical campaign (from September 2019 to January 2020) indicate that NGC 3516 is maybe awakening (Ili\'c et al. 2020). The observational facts supporting this are the increase of the continuum emission, the variability of the coronal lines, and the very broad component in the Balmer lines. It is necessary to continue the  intensive monitoring of this object to capture a possible transition phase. Moreover, using multiwavelength facilities (optical, UV, and X-ray) and different tools (spectroscopy, spectro-polarymetry) may be needed to finally describe the processes behind the changing-look phenomenon in AGNs.

\section{IMPLICATIONS TO LARGE SURVEYS}


The LSST will provide photometric light-curves of millions of AGNs, and one of its key science case is the probing of the accretion disk and BLR with the RM (Brandt et al. 2018). The key science case of MSE is the spectral RM campaign of around 5000 quasars up to the redshift of 3, providing robust estimates of time lags for the largest sample of quasars (Shen et al. 2019). In order to process such a huge amount of data, novel tools and approaches largely based on machine learning (ML) algorithms, are required. 

One challenge is how to do cross-correlation analysis of realistic light curves, which usually have problems, such as e.g. gaps, nonuniform cadence, etc. One solution could be a novel ML tool (Gaussian PROcesses for TIme Delays Estimates in AGNs - GProTIDE) for time-delays measurements that utilizes generalized Gaussian processes to model the observed light curves used for extraction of time-delays (e.g., Kova\v cevi\'c et al. 2014, 2015, 2018). Another challenge is to have codes for autonomous and fast spectral fitting of multi-component and complex quasar optical spectra. For this a solution could be a FANTASY code (Fully Automated pythoN Tool for AGN Spectra analYsis), which is currently in testing fase (see Raki\'c et al., in this proceedings). For the classification of AGNs and studies of correlation between AGN properties in multidimesional space, more advance ML techniques based on manifold learning could be used (see Jankov et al., in this proceedings). 

Finally, there is an important question about the operation strategies and survey cadences of time-domain surveys required in order to be able to extract the presented investigations (for LSST see e.g. Jones et al. 2020). For these different statistical proxies should be developed and tested on both real observed and idealized modeled light curves (Kova\v cevi\'c et al. 2021). These metrics could be used for future surveys for testing the selection of operation strategies.

\section{SUMMARY}

Here we summarize the findings of the investigations of AGNs variability done within the project "Astrophysical spectroscopy of extragalactic objects" at the Astronomical Observatory and Department of Astronomy in Belgrade. The presented results of the LoTeRM campaign of type 1 AGNs include: i) a creation of homogeneous sample of very long (10+ years) light curves, which specific feature is that they come from type 1 AGN of different spectral properties and variability, ii) a developed 2DHybrid tool for the detection of the periodic oscillations in AGNs and analysis of the underlying dynamics, iii) a capture of the changing look transition in type 1 AGNs. These findings may have an application and impact to future large surveys, such as the LSST or MSE, which key-science projects will be the investigations of quasar variability. Our group is actively contributing to the development of the LSST and MSE as members of respective Science Collaborations. Through these investigation, many tools and techniques are being developed in preparation for these and other future vast surveys.

\vskip 0.4cm

\noindent{\bf Acknowledgments.}  
This paper is devoted to Dr. Alla I. Shapovalova, Research Professor of the Special Astrophysical Observatory of the Russian Academy of Science, who sadly passed away beginning of 2019. She was a pioneer in the field of AGN spectral observation and analysis, and the great inspiration to the monitoring of AGN and studies of its variability. We acknowledge the financial support of the Ministry of Education, Science and Technological Development of the Republic of Serbia through contracts no. 451-03-68/2020-14/200104 (DI, AK, L\v CP) and 451-03-68/2020-14/20002 (L\v CP).


\references

Ahumada, R., Prieto, C.~A., Almeida, A., et al.\ 2020, ApJS, 249, 3

Afanasiev, V.~L., Popovi{\'c}, L. {\v{C}}., Shapovalova, A.~I., 2019, MNRAS, 482, 4985

Andrillat, Y., \& Souffrin, S., 1968, ApJL, 1, 111

Antonucci, R. R. J., Miller, J. S., 1985, ApJ, 297, 621

Gravity Collaboration, Amorim A., Baub{\"o}ck M., Brandner W., Bolzer M., Cl{\'e}net Y., Davies R., et al., 2021, arXiv, arXiv:2102.00068

Gravity Collaboration, Amorim A., Baub{\"o}ck M., Brandner W., Cl{\'e}net Y., Davies R., de Zeeuw P.~T., et al., 2020, A\&A, 643, A154

Bentz, M. C., Denney, K. D., Grier, C. J., et al. 2013, ApJ, 767, 149

Bentz, M.~C., Katz S., 2015, PASP, 127, 67

Bentz M.~C., Peterson B.~M., Netzer H., Pogge R.~W., Vestergaard M., 2009, ApJ, 697, 160

Blandford, R.~D., McKee, C.~F., 1982, ApJ, 255, 419

Bon, E., Jovanovi{\'c}, P., Marziani, P., et al.\ 2012, ApJ, 759, 118

Bon, E., Zucker, S., Netzer, H., et al.\ 2016, ApJS, 225, 29

Brandt, W. N., Ni, Q., Yang, G., Anderson, S. F. et al. 2018, arXiv181106542B

Cho, H., Woo, J.-H., Hodges-Kluck, E., et al., 2020, ApJ, 892, 93

Denney, K. D., De Rosa, G., Croxall, K., et al. 2014, ApJ, 796, 134

Djorgovski, S.~G., Graham M.~J., Donalek C., Mahabal A.~A., Drake A.~J., Turmon M., Fuchs T., 2016, arXiv:1601.04385

Drake, A.~J., Djorgovski S.~G., Mahabal A., et al., 2009, ApJ, 696, 870

Eisenhauer, F.\ 2019, The Very Large Telescope in 2030, 30 

Gaskell, C.~M., Sparke, L.~S., 1986, ApJ, 305, 175
 
Graham, M.~J., 2019, IAUS, 339, 147

Graham, M. J., Djorgovski, S. G., Drake, A. J., et al. 2017, MNRAS, 470, 4112

Graham, M. J., Djorgovski, S. G., Stern, D., et al. 2015, Nature, 518, 74

Hickox, R.~C., Alexander D.~M., 2018, ARA\&A, 56, 625

H{\"o}nig, S.~F., Kishimoto, M., 2017, ApJL, 838, L20

Ili\'c, D.,  Oknyansky, V.,  Popovi\'c, L. \v C., et al. 2020, A\&A,  638, 13

Ili\'c, D., Shapovalova, A. I., Popović, L. \v C. et al. 2017, FrASS, 4, 12I

Ivezi{\'c}, {\v{Z}}., Kahn, S.~M., Tyson, J.~A., et al., 2019, ApJ, 873, 111

Jones, L., Yoachim, P., Ivezic, Z., et al.\ 2020, AAS/Division for Planetary Sciences Meeting Abstracts

Joni{\'c} S., Kova{\v{c}}evi{\'c}-Doj{\v{c}}inovi{\'c} J., Ili{\'c} D., Popovi{\'c} L. {\v{C}}., 2016, Ap\&SS, 361, 101

Jovanovi{\'c}, P., Popovi{\'c}, L. {\v{C}}., Stalevski, M., et al.\ 2010, ApJ, 718, 168

Kova\v cevi\'c, A., Ili\'c, D.,  Popovi\'c, L. \v C., et al. 2021, {\it in preparation}

Kova{\v c}evi{\' c}, A.~B., P{\'e}rez-Hern{\'a}ndez, E., Popovi{\' c}, L. {\v C}., Shapovalova, A. I.,  Kollatschny, W.,  Ili{\' c}, D. \ 2018, MNRAS, 475, 2051

Kova{\v c}evi{\' c}, A.~B., Popovi{\' c}, L. {\v C}., Ili{\' c}, D. \ 2020a, Open astronomy, 29,  51

Kova{\v c}evi{\' c}, A.~B., Popovi{\' c}, L. {\v C}., Shapovalova, A. I., Ili{\' c}, D. \ 2017,  Ap\&SS, 362, 31

Kova{\v c}evi{\' c}, A.~B.,  Popovi{\' c}, L. {\v C}., Simi{\'c}, S.,  Ili{\' c}, D., 2019, ApJ, 871, 32

Kova{\v{c}}evi{\'c}, A.~B., Popovi{\'c} L. {\v{C}}., Shapovalova A.~I., Ili{\'c} D., Burenkov A.~N., Chavushyan V.~H., 2014, AdSpR, 54, 1414

Kova{\v{c}}evi{\'c}, A.~B., Popovi{\'c} L. {\v{C}}., Shapovalova A.~I., Ili{\'c} D., Burenkov A.~N., Chavushyan V.~H., 2015, JApA, 36, 475

Kova{\v{c}}evi{\'c} A.~B., Yi T., Dai X., Yang X., {\v{C}}vorovi{\'c}-Hajdinjak I., Popovi{\'c} L. {\v{C}}., 2020b, MNRAS, 494, 4069.

Li, Y.-R., Wang, J.-M., Ho, L.~C., et al.\ 2016, ApJ, 822, 4

Liu, T., Gezari, S., Coleman Miller, M. 2018, ApJL, 859, L12


MacLeod, C. L., Ross, N. P., Lawrence, A., et al. 2016, MNRAS, 457, 389

Malygin, E., Uklein, R., Shablovinskaya, E., Grokhovskaya, A., Perepelitsyn, A., 2020, CoSka, 50, 328

Mej{\'\i}a-Restrepo J.~E., Lira P., Netzer H., Trakhtenbrot B., Capellupo D.~M., 2018, NatAs, 2, 63

MSE Science Team: Babusiaux, C., Bergemann, M., Burgasser, A. et al. 2019, arXiv:1904.04907

Ofek, E.~O., Soumagnac, M., Nir, G., Gal-Yam, A., Nugent, P., Masci, F., Kulkarni, S.~R., 2020, MNRAS, 499, 5782

Oknyansky, V. L., Gaskell, C. M., Huseynov, N. A., et al. 2017, MNRAS, 467, 1496

Oknyansky, V. L., Winkler, H., Tsygankov, S. S., et al. 2019, MNRAS, 483, 558

Peterson, B.~M.\ 2001, Advanced Lectures on the Starburst-AGN, 3

Peterson, B.~M., 2004, IAUS, 222, 15

Popovi\'c, L.\v  C., 2012, NewAR, 56, 74

Popovi\'c, L. \v C., 2020, OAst, 29, 1P

Popovi{\'c}, L. {\v{C}}., Kova{\v{c}}evi{\'c}-Doj{\v{c}}inovi{\'c} J., Mar{\v{c}}eta-Mandi{\'c} S., 2019, MNRAS, 484, 3180

Popovi\'c, L.\v C., Shapovalova, A.I., Ili\'c, D. et al., 2011, A\&A, 528, 130

Popovi\'c, L.\v C., Simi\'c, S., Kova{\v{c}}evi{\'c} A.~B., Ili\'c, D., 2021, MNRAS, {\it submitted}

Runco, J. N., Cosens, M., Bennert, V. N., et al. 2016, ApJ, 821, 33

Shapovalova, A. I., Burenkov, A. N., Carrasco, et al., 2001, A\&A, 376, 775

Shapovalova, A. I., Doroshenko, V. T., Bochkarev, et al., 2004, A\&A, 422, 925

Shapovalova, A.I., Popovi\'c, L. \v C., Afanasiev, V.L, 2019, MNRAS, 485, 4790

Shapovalova, A. I.; Popovi\'c, L. \v C., Bochkarev, N. G., 2009, NewAR, 53, 191S

Shapovalova, A. I., Popovi\'c, L. \v C., Burenkov, A. N., et al., 2010a, A\&A, 517, 42 

Shapovalova, A. I., Popovi\'c, L. \v C., Burenkov, A. N., et al., 2013, A\&A, 559, 10

Shapovalova, A. I., Popovi\'c, L. \v C., Burenkov, A. N., et al., 2010b, A\&A, 509, 106

Shapovalova, A. I., Popovi\'c, L. \v C., Burenkov,  et al., 2012, A\&ASS, 202, 10

Shapovalova, A. I., Popovi\'c, L. \v C, Chavushyan, et al., 2017, MNRAS, 466, 4759

Shapovalova, A. I., Popovi\'c, L. \v C., Chavushyan, V. H., et al., 2016, ApJS, 222, 25

Shapovalova, A. I., Popovi\'c, L. \v C., Collin, S., et al., 2008, A\&A, 486, 99

Shen, Y., Anderson, S., Berger, E., et al., 2019, Bulletin of the American Astronomical Society, 51, 3, 274, Science White Paper for the US Astro 2020 Decadal Survey

Shen Y., Grier C.~J., Horne K., Brandt W.~N., Trump J.~R., Hall P.~B., Kinemuchi K., et al., 2019, ApJL, 883, L14

Simi{\'c} S., Popovi{\'c} L. {\v{C}}., 2016, Ap\&SS, 361, 59

Songsheng, Y.-Y., Li, Y.-R., Du, P., et al.\ 2021, arXiv:2103.00138

Stalevski, M., Asmus, D., \& Tristram, K. R. W. 2017, MNRAS, 472, 3854

Stalevski, M., Tristram, K. R. W., \& Asmus, D. 2019, MNRAS, 484, 3334

Gravity Collaboration, Sturm E., Dexter J., Pfuhl O., Stock M.~R., Davies R.~I., Lutz D., et al., 2018, Nature, 563, 657

Sulentic, J. W., Marziani, P., Dultzin-Hacyan, D. 2000, ARA\&A, 38, 521

Urry, C. M., Padovani, P., 1995, PASP, 107, 803

Wang J.-M., Bon E., 2020, A\&A, 643, L9

Wang J.-M., Du P., Brotherton M.~S., Hu C., Songsheng Y.-Y., Li Y.-R., Shi Y., et al., 2017, NatAs, 1, 775

Wang, J.-M., Songsheng, Y.-Y., Li, Y.-R., et al.\ 2020, NatAs, 4, 517

Woo J.-H., Cho H., Gallo E., Hodges-Kluck E., Le H.~A.~N., Shin J., Son D., et al., 2019, NatAs, 3, 755

\endreferences

\end{document}